\documentclass[11pt,twoside]{article}
\usepackage{asp2010}

\resetcounters

\begin{document}

\title{The Andromeda Optical and Infrared Disk Survey}
\author{Jonathan~Sick,$^1$ St\'{e}phane~Courteau,$^1$ and Jean-Charles~Cuillandre$^2$}
\affil{$^1$Queen's University, Kingston, Ontario Canada K7L 3N6}
\affil{$^2$Canada-France-Hawaii Telescope Corporation, Kamuela, HI 96743, USA}

\begin{abstract}
The Andromeda Optical and Infrared Disk Survey has mapped M31 in $u^* g^\prime r^\prime i^\prime J K_s$ wavelengths out to $R=40$~kpc using the MegaCam and WIRCam wide-field cameras on the Canada-France-Hawaii Telescope.
Our survey is uniquely designed to simultaneously resolve stars while also carefully reproducing the surface brightness of M31, allowing us to study M31's global structure in the context of both resolved stellar populations and spectral energy distributions.
We use the Elixir-LSB method to calibrate the optical $u^* g^\prime r^\prime i^\prime$ images by building real-time maps of the sky background with sky-target nodding.
These maps are stable to $\mu_g \lesssim 28.5$~mag~arcsec$^{-2}$ and reveal warps in the outer M31 disk in surface brightness.
The equivalent WIRCam mapping in the near-infrared uses a combination of sky-target nodding and image-to-image sky offset optimization to produce stable surface brightnesses.
This study enables a detailed analysis of the systematics of spectral energy distribution fitting with near-infrared bands where asymptotic giant branch stars impose a significant, but ill-constrained, contribution to the near-infrared light of a galaxy.
Here we present panchromatic surface brightness maps and initial results from our near-infrared resolved stellar~catalog.
\end{abstract}

\section{Introduction to the ANDROIDS Project}

The Andromeda Galaxy (M31) is a special laboratory for testing our understanding of galaxy formation and evolution.
Its close proximity enables detailed mapping of the star formation histories imprinted in resolved stellar populations, while our external vantage point permits detailed decompositions of galaxy structures.
Here we present the Andromeda Optical and Infrared Disk Survey (ANDROIDS): a homogeneous mapping of the entire Andromeda Galaxy from near-UV to near-infrared (NIR) wavelengths with imaging that simultaneously resolves stars and accurately recovers surface brightness.
This survey is being carried out with the MegaCam ($u^*g^\prime r^\prime i^\prime$) and WIRCam ($JK_s$) instruments on the Canada-France-Hawaii Telescope.
ANDROIDS improves upon previous all-disk star catalogs \citep[e.g., the Local Group Galaxy Survey;][]{Massey:2006} with sharper seeing, superior $u^*$ sensitivity and incorporation of deep NIR bands.
Previous surface brightness investigations of the M31 disk have either barely reached the 10~kpc star forming ring in SDSS optical imaging \citep{Tamm:2012} or the bulge in 2MASS NIR imaging \cite{Beaton:2007}.
The ANDROIDS survey uses rigorous background subtraction strategies to obtain robust surface brightness measurements out to $R=40$~kpc along the disk.

A prime motivation for ANDROIDS, besides mapping M31's structure and stellar content, is to understand the systematics of stellar population synthesis, particularly in the NIR.
Combining NIR and optical data is invaluable for alleviating the well-known degeneracy between the stellar metallicity, age and dust content on a galaxy's colors.
However, the wide-spread use of NIR bands has been stunted by our inability to consistently include them in spectral energy distribution (SED) fits, as shown nicely by \cite{Taylor:2011}.
This failure may be due either to inadequate modelling of star formation histories and dust, or because the NIR light itself is dominated by asymptotic giant branch (AGB) stars whose light is a challenge to most population synthesis models.
The ANDROIDS survey is well suited to identify and resolve tensions in NIR stellar population synthesis since a region's SED can be measured, while the resolved stars that contribute to that SED are also measured.

\section{Low Surface Brightness Imaging of M31}

\begin{figure}[t]
\centering
\includegraphics[width=\columnwidth]{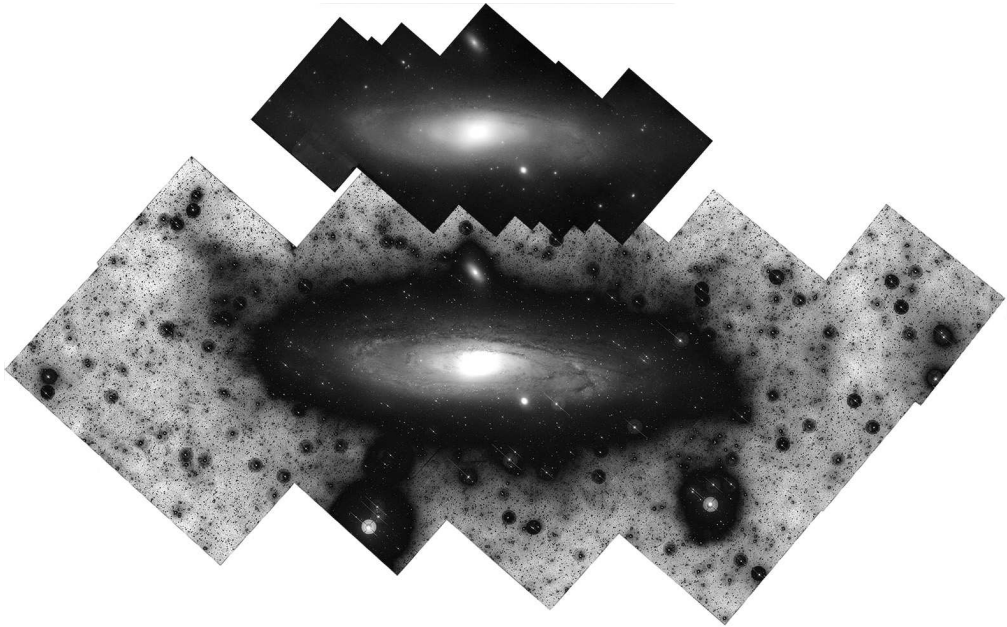}
\caption{ANDROIDS $J$-band mosaic (top) and MegaCam mosaics (bottom). The optical mosaics are stable down to $\mu_g<28.5$~mag~arcsec$^{-2}$, and reveal outer disk features such as the northern spur.}
\label{fig:mosaics}
\end{figure}

Covering the Andromeda Galaxy's disk requires fourteen 1-square degree pointings with MegaCam, but the background in these fields cannot be directly measured.
For our CFHT/MegaCam optical mapping we have used a new method called Elixir-LSB that enables calibrated surface brightness maps via real-time background monitoring  using a sky-target-target-sky nodding pattern.
This yields optical mosaics with modest integrations (15~minutes in $g^\prime r^\prime i^\prime$, 45~minutes in $u^*$) that reach surface brightness levels of $\mu_g \lesssim 28.5$~mag~arcsec$^{-2}$, while also having a pixel scale of 0.7~pc on the disk of M31.
The optical mosaics (Fig.~\ref{fig:mosaics}) have stable surface brightness calibration out to the mosaic edge at $R=40$~kpc, and for the first time, the northern spur in the M31 disk is seen in \emph{integrated} light, after first being detected by \cite{Ferguson:2002} in stellar density maps.

Producing equivalent surface brightness maps at $J$ and $K_s$ is far more difficult, as outlined by \cite{Sick:2013}.
NIR sky glow is both bright---$10^3\times$ M31's NIR surface brightness at $R=20$~kpc---and variable in space and time.
\citeauthor{Sick:2013} find that nodding across a target as large as M31 introduces background level uncertainties at a level of 2\% of the NIR background.
However we can produce NIR mosaics (Fig.~\ref{fig:mosaics}) stable to $<0.1$\% of the background level by solving for a set of background level offsets for each image that minimize image-to-image surface brightness discontinuities.
Since the background corrections are so large, absolute surface brightness must be calibrated by appealing to star counts.

\section{Near-Infrared Stellar Populations}

\begin{figure}[t]
\centering
\includegraphics[width=\columnwidth]{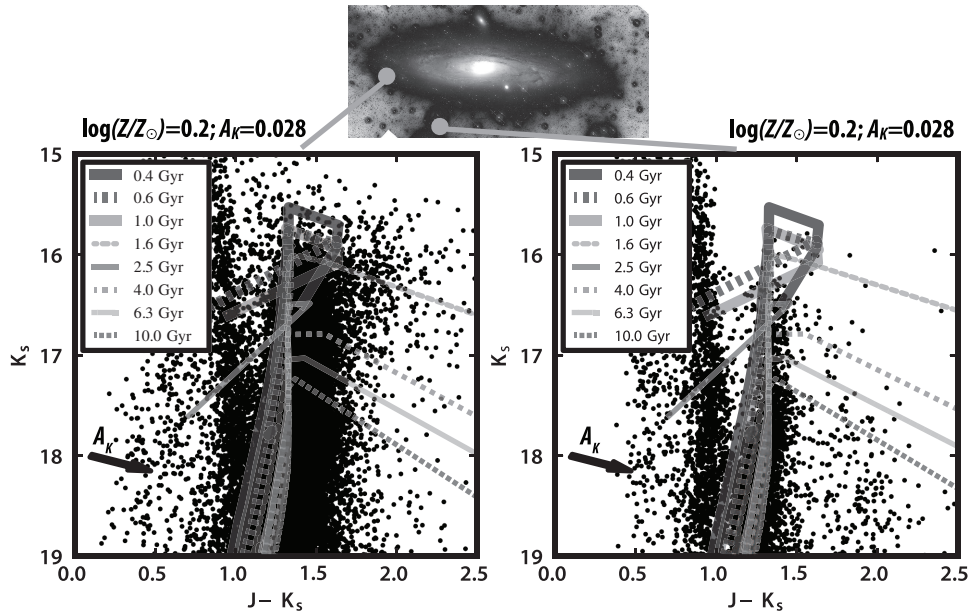}
\caption{Near-infrared color magnitude diagrams with $\log Z/Z_\odot=0.2$ metallicity Padova isochrones overlaid for various ages.
Tight RGB sequences in the disk-halo transition (right) are nearly mono-abundance, while those in the outer optical disk (left) are broader due to multiple stellar populations and are redder than expected, likely due to additional extinction.}
\label{fig:cmd}
\end{figure}

Our ANDROIDS NIR color magnitude diagrams are sensitive to AGB and upper red giant branch (RGB) stars in M31.
A great advantage of NIR over optical photometry is that M31 NIR RGB/AGB sequences have distinct colors compared those of the Milky Way foreground.
Indeed, this is why previous resolved stellar population analyses of M31 \citep[e.g.,][]{Williams:2003} have been restricted to young ($<500$~Myr) populations.
In Fig.~\ref{fig:cmd} we plot high metallicity ($\log Z/Z_\odot = 0.2$) Padova isochrones \citep{Marigo:2008} against two $J-K_s$ CMDs in the outer M31 disk.
We find that our outer-disk NIR stellar populations are consistent with predominantly intermediate-age ($>1$~Gyr) high-metallicity population in agreement with HST observations by \cite{Brown:2006}.

\section{Future Work}

ANDROIDS is synergistic with the high-resolution, but spatially-limited, Panchromatic Hubble Andromeda Treasury Survey \citep{Dalcanton:2012}, and the extended halo star counts by the Pan-Andromeda Archaeological Survey \citep{McConnachie:2009}.
We will combine these surveys with ANDROIDS to fully characterize the stellar mass distribution of the bulge, disk and halo \citep[e.g.,][]{Courteau:2011}, and map star formation histories and dust content in M31.
By confronting SED fitting with resolved stellar populations, we will identify the tensions in NIR stellar population synthesis.

\acknowledgements We thank Marc Seigar for organizing a wonderful conference. JS and SC acknowledge the support of NSERC.

\bibliography{sickj}

\end{document}